\def\year{2022}\relax
\documentclass[letterpaper]{article} 
\usepackage{aaai22}  
\usepackage{times}  
\usepackage{helvet}  
\usepackage{courier}  
\usepackage[hyphens]{url}  
\usepackage{graphicx} 
\usepackage{tabularx}
\usepackage{xcolor}
\definecolor{mit}{HTML}{A31F34}
\usepackage{natbib}  
\usepackage{caption} 
\DeclareCaptionStyle{ruled}{labelfont=normalfont,labelsep=colon,strut=off} 
\usepackage{algorithm}
\usepackage{algorithmic}
\usepackage{newfloat}
\usepackage{listings}
\usepackage{amsmath}
\usepackage{makecell}

\usepackage{lipsum}
\usepackage{array}

\floatstyle{ruled}
\newfloat{listing}{tb}{lst}{}
\floatname{listing}{Listing}
\title{Introducing Variational Autoencoders to High School Students}
\author {
    Zhuoyue Lyu,\textsuperscript{\rm 1}
    Safinah Ali, \textsuperscript{\rm 2}
    Cynthia Breazeal \textsuperscript{\rm 2}
}
\affiliations {
    \textsuperscript{\rm 1} University of Toronto, 27 King's College Circle, Toronto, Ontario, Canada, M5S 1A1\\
    \textsuperscript{\rm 2} MIT Media Lab, 75 Amherst Street, Cambridge, Massachusetts, United States, 02139\\
    zhuoyue@cs.toronto.edu, \{safinah, cynthiab\}@media.mit.edu
}

\begin{document}
\maketitle

\begin{abstract}
Generative Artificial Intelligence (AI) models are a compelling way to introduce K-12 students to AI education using an artistic medium, and hence have drawn attention from K-12 AI educators. Previous Creative AI curricula mainly focus on Generative Adversarial Networks (GANs) while paying less attention to Autoregressive Models, Variational Autoencoders (VAEs), or other generative models, which have since become common in the field of generative AI. VAEs' latent-space structure and interpolation ability could effectively ground the interdisciplinary learning of AI, creative arts, and philosophy. Thus, we designed a lesson to teach high school students about VAEs. We developed a web-based game and used Plato's cave, a philosophical metaphor, to introduce how VAEs work. We used a Google Colab notebook for students to re-train VAEs with their hand-written digits to consolidate their understandings. Finally, we guided the exploration of creative VAE tools such as SketchRNN and MusicVAE to draw the connection between what they learned and real-world applications. This paper describes the lesson design and shares insights from the pilot studies with 22 students. We found that our approach was effective in teaching students about a novel AI concept.
\end{abstract}



\section{Introduction}
Generative Artificial Intelligence (AI) models have been integrated into the K-12 curricula, but the focus is mainly on Generative Adversarial Networks (GANs)~\cite{goodfellow2014generative} with little attention on Autoregressive Models, Variational Autoencoders (VAEs)~\cite{kingma2013auto} or other generative models. The continuity of VAE's latent space enables interpolation, which provides a creative way to explore artistic ideas. Hence, it is no surprise that VAEs are increasingly being used for media generation like creating music~\cite{MusicVAE}, sketches~\cite{ha2017neural}, or 3D shapes~\cite{pointflow}. In this work, we explore ways to teach high school students about VAEs and their applications, with the aim for these approaches to be included in high school AI curricula.

We found that the representation captured by VAE's latent space is similar to the \textit{form} in Plato's theory of forms~\cite{sedley_2016} where Plato argues that all we see in this world are merely the shadows of perfect ideas (forms). Plato illustrated this idea using the cave allegory~\cite{platoCave}. Similarly, VAE's latent space (forms) is the compressed representation of high-dimensional data (shadows), which could be used to reconstruct new outputs (new shadows). Inspired from previous work that suggested students demonstrated significant interests in learning AI using Philosophical approaches~\cite{ellis2004teaching, ellis2005organizing}, we used Plato's cave to introduce how VAEs work.

 K-12 AI lessons have successfully utilized interactive tools that help students understand complex AI algorithms. For instance, \citet{wan2020smileycluster}~developed a learning environment to teach students k-means clustering. \citet{ali2021gans}~used a collaborative web game to teach the inner workings of GANs. \citet{lee2021contour}~utilized an interactive web game to teach how neural networks work. Motivated by the success of these approaches, we developed a shadow matching game to facilitate understandings of the encoder and decoder in a VAE and a Colab notebook to let students re-train VAEs with hand-written digits.
 
Findings from pilot studies with 22 participants indicate students understood the role of the encoder and decoder after the lesson and enjoyed the hands-on experience of re-training VAEs and playing with real-world applications. This interdisciplinary lesson could either be taught stand-along, combined with ``What are GANs?"  lesson~\cite{ali2021gans} into the existing Creative AI curricula~\cite{ ali2021exploring, ali2019constructionism}, or as a sub-module for the other AI and Philosophy curricula.

\section{Background}
We developed a high school AI lesson focusing on teaching VAEs through creative applications, interactive tools, and a philosophical metaphor. 

\subsection{Creative AI Education}
There are existing Creative AI education approaches that focus on the technical constituents and applications of GANs, as well as ethical implications such as Deepfakes~\cite{virtue2021gans, williams2019artificial, ali2021gans, ali2021exploring,ALI2021100040}. While these approaches successfully taught students how generative models work through creative applications, they focused on GANs, which are not the only generative AI models. In this work, we focus on VAEs~\cite{kingma2013auto}, another powerful generative model extensively used to represent high-dimensional data via a low-dimensional latent space~\cite{girin2020dynamical}. Inspired by previous work, students' learning is guided through playing a simulated Shadow Matching Game that teaches students the constituents of a VAE, and an exploration with tools that use VAEs to create media. Previous work mentions the limitation that while students interacted with GANs' tools, they did not train a network with custom datasets and see it in action~\cite{ali2021exploring}. In this work, we made use of a scaffolded Colab notebook, Digits Interpolation Notebook, where students re-train VAEs on a hand-written digits dataset. Both the game and the notebook are outlined in the Software Tools section.

\subsection{Variational Autoencoders (VAEs)}
VAE is a special kind of Autoencoder (AE). An AE consists of an encoder, latent space, and a decoder, where the latent space tends to be small (bottleneck) to capture the critical information from the inputs. A VAE's latent space differs from an AE's as its encoder produces two vectors: means and variances that define distributions in the latent space, and its decoder randomly samples from distributions to reconstruct the outputs. This provides continuity to the discrete latent spaces of the AEs', enabling the interpolation and morphing among samples. VAEs lessons are common for college-level students~\cite{MIT6S191, stanfordCS236, uoftCSC412}, but those require mathematics and statistics foundations that K-12 students do not have.  In our lesson, we focused on intuition and applications instead. VAEs as a subgroup of generative models have demonstrated creativity in music, drawings, and 3D objects generation: MusicVAE~\cite{MusicVAE} was trained on melodies and beats, its web-based tools Melody Blender and Beats Mixer let users explore melodies and beats interpolation. SketchRNN~\cite{ha2017neural} is able to complete the users' sketches, performing interpolations, and mimic users' drawings. PointFlow~\cite{pointflow} can interpolate between objects constructed by 3D points.

\subsection{AI and Philosophy}
Philosophical concepts have been used to teach AI and vice versa. For example, when Ellis and Andam taught machine consciousness to high school students using the Turing test~\cite{Turing2009}, they noticed it was the philosophical content that interested students the most~\cite{ellis2004teaching}. \citet{ellis2005organizing} then presented a concept map for K-12 around AI and Philosophy, focusing on Decartes' Mind and Body~\cite{radner1971descartes} as well as the philosophy of mind, intelligence, and consciousness as a way to supports a deeper understanding of AI and the philosophical issues surrounding that. Sloman proposed teaching computing as a way to do philosophy~\cite{sloman2009teaching}, Lim similarly proposed using machine learning for teaching induction~\cite{lim2020philosophy}. Given that the representation captured by the VAE's latent space is similar to Plato's theory of forms~\cite{sedley_2016}, we used Plato's cave~\cite{platoCave}, an ancient philosophical allegory that discussed the problem of reality and knowledge, to introduce how VAEs work.

\begin{figure}[htb!]
\centering
\includegraphics[width=0.9\columnwidth]{./pic/platoCave.jpeg}
\caption{The Plato's Cave ~\cite{cavePic, platoCave}}
\end{figure}

\section{Software Tools}
To facilitate the learning objectives, we designed two web-based tools: Shadow Matching Game and Digits Interpolation Notebook to help students understand the role of a VAE's encoder, latent space, and decoder, and enabling the re-training of VAEs using their hand-written digits.

\subsection{Shadow Matching Game}
 This game was developed in Unity~\cite{unity}. The original version was designed in a Virtual Reality (VR) environment, and the user can grab the cubes and move them around using their virtual hands and rotate the object using controllers. However, due to the pandemic and limited access to VR headsets, we adapted the activity for a web browser using WebGL.
 

\begin{figure}[htb!]
\centering
\includegraphics[width=\columnwidth]{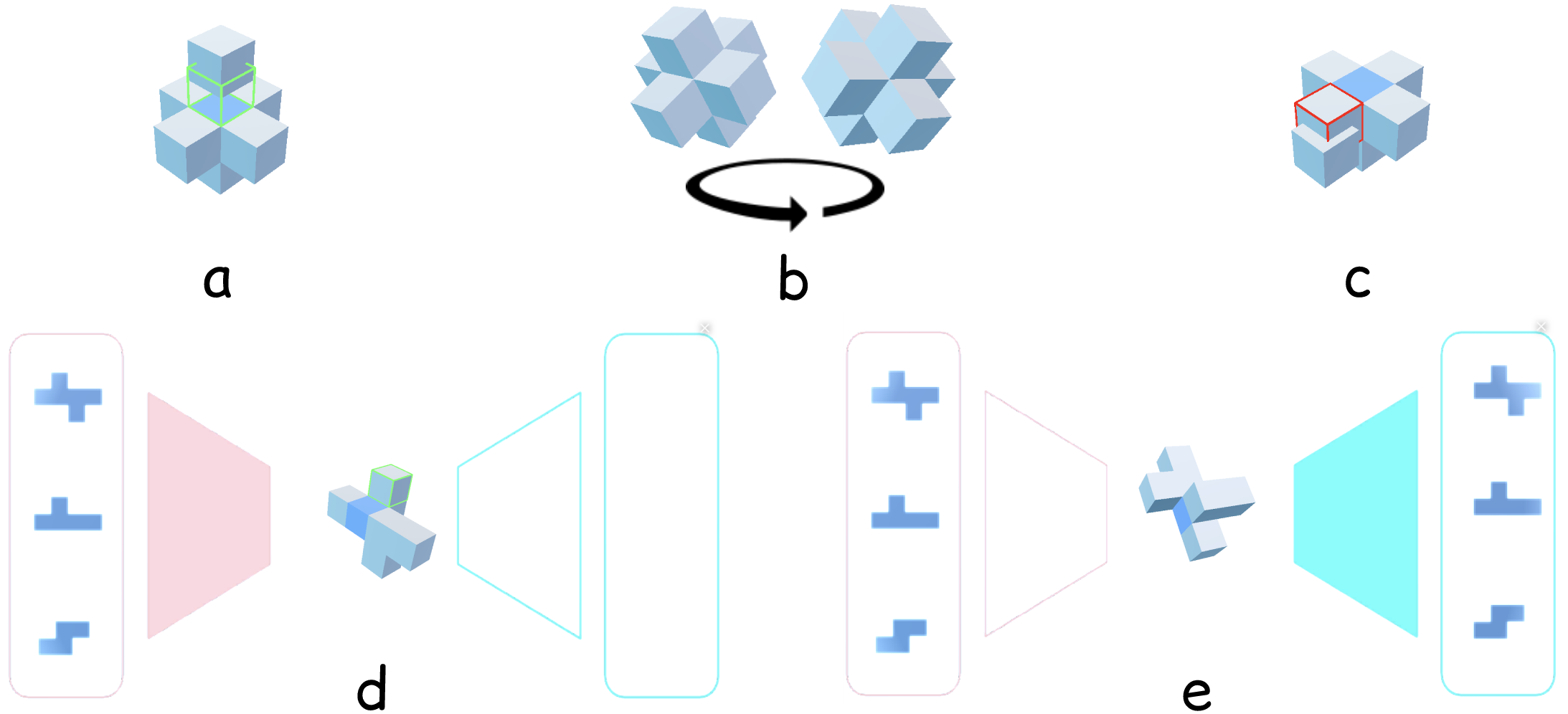}
\caption{The Shadow Matching Game: (a) target position available; (b) rotating the object using arrow keys; (c) target position not available; (d) encoder mode; (e) decoder mode.}
\label{shadowMatching}
\end{figure}


\subsubsection{Drag and Rotate}
The object in the middle represents latent space, and it is constructed by multiple cubes. The user can drag each cube using the cursor, and the cube will snap to the closet position where its X, Y, Z coordinates are integers. An outline shows up when dragging, indicating if the destination is available (green) as in Figure~\ref{shadowMatching} (a) or not (red) as in Figure~\ref{shadowMatching} (c). The object that represents latent space can be rotated using arrow keys on the keyboard as in Figure~\ref{shadowMatching} (b). The ``S" key on the keyboard is used to snap the object's rotation to the closest angle of the multiple of 90 degrees.

\subsubsection{Encoder and Decoder}
The game starts with encoder mode as in Figure~\ref{shadowMatching} (d), and the user is free from moving the cubes around. By pressing the key ``D", the user enters the decoder mode, as in Figure~\ref{shadowMatching} (e) where the user can create shadows. However, the user can not move cubes in the decoder mode, thus should press ``E" to return to encoder mode to do that, but this will erase all output shadows.

\subsubsection{Stop and Next}
The timer will stop when the user creates three shadows using the decoder, and it will resume when the user goes back to the encoder mode. The game starts with \textit{Easy} that has 7 cubes, but the user can progress to the next level, \textit{Hard} with 27 cubes, by pressing ``N" key.

\subsubsection{VAE and AE}
\label{VAEAE}
The current game version represents an Autoencoder (AE) but aids understanding of a VAE. For the VAE version, instead of updating one object, the user is given three objects in the latent space, and the user needs to update all three objects to match a specific shadow. When generating new shadows, the decoder would randomly pick one of three objects to cast a shadow. The ``three objects" represent the distributions generated by VAE's encoder, while the ``randomly pick" represents the decoder's random sampling process when reconstructing outputs.

\subsection{Digits Interpolation Notebook}
We developed this tool in Google Colab based on the existing VAE notebook~\cite{10.1145/3277644.3277775}, which was designed for Machine Learning experts. We simplified it with the UI enabled by Colab's Forms feature~\cite{colabForms} such that users with zero coding experience can still re-train their VAEs. In addition, we adapted Drawing on Colab~\cite{colabDraw} to generate data within the notebook by letting the user draw custom digits. Finally, we added the algorithm to handle the custom dataset, animation (GIF) of the interpolation, and visualization of inputs and outputs.

\begin{figure}[htb!]
\centering
\includegraphics[width=\columnwidth]{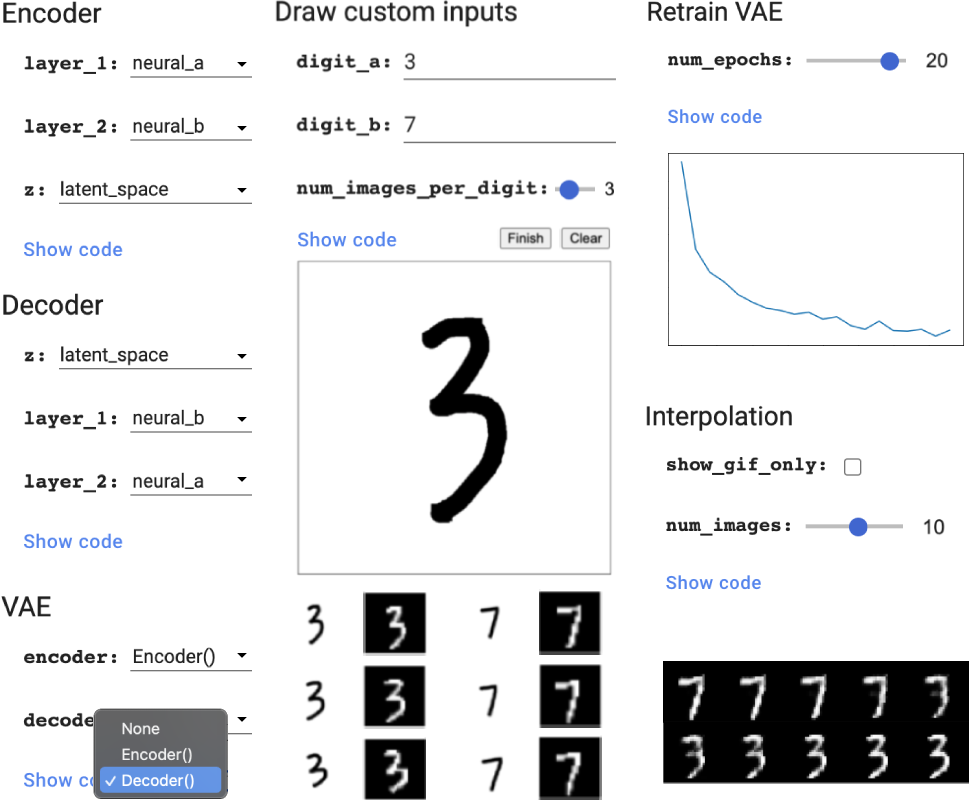}
\caption{The Digits Interpolation Notebook}.
\label{notebookVis}
\end{figure}

\subsubsection{Preparing a VAE}
The first ``Setup" block will import prerequisite libraries. The following two blocks allow users to select among {\small\texttt{neural\_a}}, {\small\texttt{neural\_b}}, and {\small\texttt{latent\_space}} for the respective layers. The user can select the correct option based on the visual at the beginning of the notebook. {\small\texttt{Layer\_1}} of the encoder is different from that of the decoder. Thus the user should select a different neural layer.

\subsubsection{Dataset and Re-training}
The ``Draw custom inputs" block allows users to define the endpoints of interpolation using {\small\texttt{digit\_a}} and {\small\texttt{digit\_b}}. {\small\texttt{num\_images\_per\_digit}} defines the number of drawings, enables the user to explore the relationship between the size of the dataset and output quality. A canvas would pop up for the user to draw the digits. The ``Finish" button records the current drawing and proceed to the next, while the ``clear" would erase it. After the drawing is done, a picture indicating the training set and test set shows up. The user can now re-train VAE on those digits.

\subsubsection{Interpolation and More}
The interpolation section creates a GIF animation based on the images that VAE generates. The user can choose to see those individual images by unchecking the {\small\texttt{show\_gif\_only}}. The {\small\texttt{num\_images}} defines the number of samples between the two interpolation endpoints, thus determines the smoothness of the animation. The ``Advanced" section allows experienced students to explore the latent space, reconstruction, and 2D interpolation.

\section{Lesson Design}
\label{Curriculum}

\begin{figure*}[t]
\centering
\includegraphics[width=\textwidth]{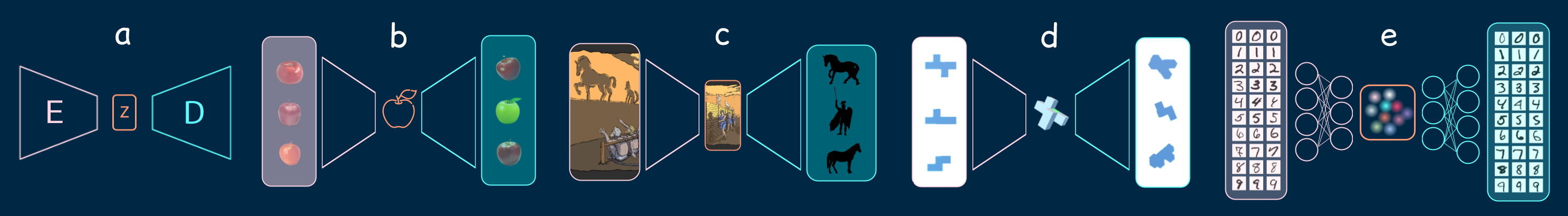}
\caption{Five different illustrations of the VAEs used throughout the lesson: (a) VAE; (b) VAE with Plato’s Theory of Forms; (c) VAE with Plato's Cave; (d) VAE with Shadow Matching Game; (e) VAE with Digits Interpolation Notebook.}
\label{VAE}
\end{figure*}

In this section, we first describe the setup and targets of our VAEs lesson and then present the details of all three modules. The lesson took a maximum of 120 minutes with 30-40 minutes for each module.

\subsubsection{Setup and Resources Needed}
The lesson requires a laptop that runs a modern web browser with a valid Google account. All resources are available at {\textbf{{raise.mit.edu/vae}}}


\subsubsection{Target Age Group}
This lesson is designed for high school students, but the user study showed positive feedback from middle school students as well. Modules were designed with the flexibility to make them adaptable. For instance, the shadow matching game has two levels of difficulties, the Colab notebook can be used with UI (easy) or with actual Python code (hard), and the Plato's cave can be taught with only the allegory (easy) or with Plato's theory of forms (hard). Some experiences in AI, neural networks, and programming will be helpful.

%

\subsubsection{AI Concepts Addressed}
This lesson targets VAEs, a kind of Generative AI model. It explains the structure of VAEs as the encoder, latent space, and decoder and explores deeper into the philosophical and artistic values of the interpolation and latent space. It also touches upon AI concepts such as neural networks, training/test set, accuracy/loss, and underfit/overfit through the simplified machine learning development cycle in the Colab notebook activity.

 \subsubsection{Expected Learning Outcomes}
At the end of the lesson, students will be able to: (1) Describe the roles of the encoder and the decoder; (2) Describe the structure of the latent space and how interpolation is performed; (3) Given a VAE, identify what the latent space, encoder, and the decoder is; (4) Given a VAE, identify what the inputs and outputs to the encoder and decoder are.

\subsection{Module 1: What are VAEs?}
This module aims to provide a basic introduction to VAEs. It starts with examples of creative arts by humans and AIs to trigger students' curiosity. Then, through Shadows and Plato's Cave, students begin to understand the essence of VAEs. The module ends with a game to help students understand the roles of the encoder, decoder, and latent space.

\subsubsection{Activity 1: Human or AI?}
Researchers pointed out that it is essential to engage students' initial understanding~\cite{bransford2000people} in teaching novel concepts. We provide three media for learners to categorize as created by human or AI: a human-created sketch, a sketches generated by the SetchRNN, a melodies interpolation created by MusicVAE.

\subsubsection{Activity 2: Shadows and Plato's Cave}
\label{M1A2}
We show a picture of a cylinder projecting two different shadows to suggest there is an underlying \textit{truth} among \textit{true} appearances, then introduce Plato's cave~\cite{platoCave}. The shadows and Plato's Cave analogy suggest that the same object could create multiple shadows or appearances, which is precisely the essence of VAEs.

\subsubsection{Activity 3: Shadow Matching Game}
\label{M1A3}
The game connects the shadows and Plato's cave analogy to the VAE by letting the students play the roles of the encoder and decoder: The object in the middle represents the latent space, reshaping objects to match shadows represents the role of the encoder and creating shadows by rotating the object represents the role of the decoder. It is a role-playing game to build intuition, and the students were not required to write any code. 



\subsection{Module 2: Building a VAE}
This module dives deeper into the structure of VAEs and provides hands-on experience in training AI. The actual structure of the VAEs is revealed and touches upon the concepts of neural networks and distributions. Students then look at how interpolation is performed on the latent space and finally re-train their own VAEs using their custom digits on the notebook we designed.

\subsubsection{Activity 1: Real VAEs?}
We reveal the structure of a VAE and let students guess what the encoder and decoder are made of (artificial neural networks) and what latent space is (distributions). We briefly mention that neural networks can take information from the previous layer and pass it to the next. The clusters (distributions) in the latent space represent how many categories the encoder found in the inputs. 

\subsubsection{Activity 2: Look, Interpolation!}
Students are shown to the MNIST dataset~\cite{lecun-mnisthandwrittendigit-2010} that consists of 60,000 written digits, with a question: will VAEs performs well with MNIST? By drawing the line between two points on the latent space, we can interpolate between them. Students would recall the sketches and melodies that they saw in the first activity of the previous module, and the teacher should emphasize that that is because they are all VAEs and thus, can perform the interpolation.

\subsubsection{Activity 3: Digits Interpolation Notebook}
\label{M2A3}
This notebook allows students to apply what they learned and re-train VAEs to perform interpolations on their hand-written digits. This activity provides students with the experience of the complete cycle of Machine Learning development in a simplified setting: building models, providing datasets, training networks, tuning hyper-parameters, and checking outcomes.

\begin{figure}[htb!]
\centering
\includegraphics[width=0.9\columnwidth]{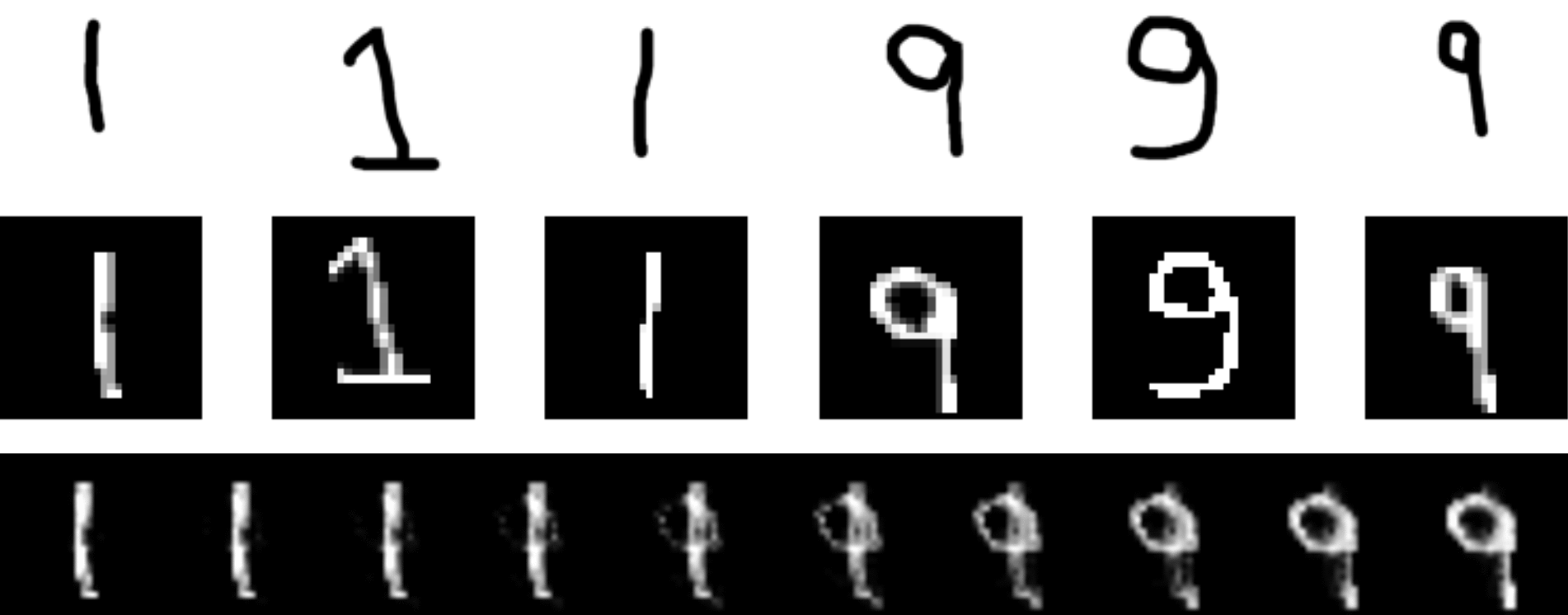}
\caption{Example work collected from a student using the Digits Interpolation Notebook in the user study. The student drew the digits using the mouse ($1^{\text{st}}$ row), algorithm processed the inputs to 28 x 28 ($2^{\text{nd}}$ row), and the student re-trained VAEs to perform interpolation ($3^\text{rd}$ row).}
\label{studentInterpolaiton}
\end{figure}




\subsection{Module 3: Exploring VAEs}
This module consolidates students' understanding of VAEs. Students first explore tools built with modern VAE models and identify encoder, decoder, latent space. Then, students take a quiz to recall what they have learned. The lesson concludes with the roadmap of their learning and guides a artful and philosophical discussion surrounding VAEs. 



\subsubsection{Activity 1: VAEs and Arts?}
\label{M3A1}
Students are introduced to the Interpolation and Mimic Drawings tools by SketchRNN. They are expected to find that sketches quality worsens as they increase the \textit{temperature}. Then, students play with Melody Mixer and Beat Blender to explore the music interpolation with questions ``How does the bar numbers in the Melody Mixer affect interpolation?" ``Where is the palette of the Beat Blender?" Since students have seen these VAEs in Module 1, this helps them build a more robust connection.


\subsubsection{Activity 2: Loss \& Accuracy!}
\label{M3A2}
To test students' understanding of the course materials, we designed a quiz with five questions directly related to the materials they have learned throughout the course using Kahoot!~\cite{dellos2015kahoot}, an online game-based learning platform. The name ``Loss \& Accuracy!" came from the AI training: students are AI models, it is time to check their learning progress. The questions and options of the quiz are in the Results section.

\subsubsection{Activity 3: Our VAEs Journey}
\label{M3A3}
The session ends with a road map with four key activities: (1) Shadows and Plato's Cave; (2) Shadow Matching Game; (3) Digits Interpolation Notebook; (4) VAEs and Arts. Students are encouraged to draw connections among those activities: Plato's cave suggests there is an underlying truth among shadows, which is the essence of VAEs since the encoder figures out the truth while the decoder reconstructs shadows from it; the unique structure of the latent space enables digits interpolation and sketch/melody progression. The lesson is concluded with a survey that collects students' feedback towards activities and the lesson as a whole. 

\section{User Study}

\subsection{Recruitment}

Participants were recruited via a call for participation we sent to public schools' mailing lists. Participants were informed that the pilot session would be recorded and their feedback would be used to improvise the curriculum. We received 38 responses, out of which we arranged 11 sessions in total based on the participants' and instructors' availabilities. All participants and their parents signed the assent and consent forms respectively for participating in the study and for recording the video and audio of the sessions. 

\subsection{Participants}
22 participants (F = 12, M = 10) joined the study. Most participants were high-schoolers (grades 9 through 12), with a few exceptions: one from grade 7, three from grade 8, and one rising junior undergraduate student. Each session was led by one teacher, with 1-4 participants. The average duration of the sessions was 1 hour 45 minutes. Due to the pandemic, all sessions were conducted online through Zoom~\cite{zoom}.

\subsection{Assessment}
We collected students' responses in the ``Human or AI?" activity from Module 1,  ``Loss \& Accuracy!" and ``Our VAEs Journey" activities from Module 3. We also collected students' interpolation work (Figure~\ref{studentInterpolaiton}) in the Digits  Interpolation  Notebook and comments throughout the session. 
\section{Results}

\subsection{Pre-assessment}
\begin{figure}[t]
\centering
\includegraphics[width=0.98\columnwidth]{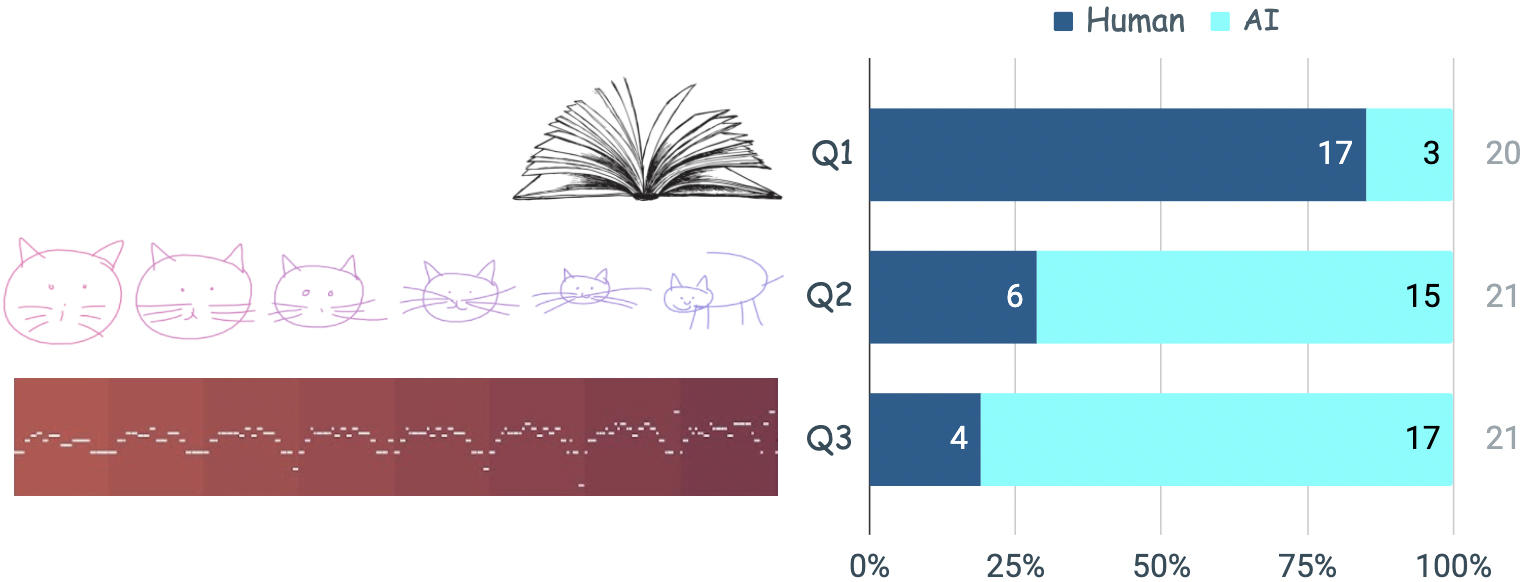} 
\caption{Responses breakdown for ``Human or AI?"}.
\label{M1A1-fig}
\end{figure}

Students completed this assessment in the ``Human or AI?" activity of Module 1. Out of 22 participants, 20 valid responses are collected for Q1, 21 for both Q2 and Q3. Among them, 17 students  (85\%) correctly selected ``Human" for Q1, 15 students  (71.4\%) and 17 students (81.0\%)  correctly chose ``AI" for the Q2 and Q3, respectively (Figure~\ref{M1A1-fig}).

\subsubsection{Q1. Human-created Sketch} Some students believed that AI created it:
\begin{quote}
\textit{    I think the drawing method looks very standardized, and despite having a more ``rustic" feel to it, the drawing still looks very even.
}\end{quote}
However, the majority of them believed it is human-created:
\begin{quote}
\textit{ The shading seems to be a humans attempt to mimic actual shades ... a machine would be more accurate.}
\end{quote}

\subsubsection{Q2. SketchRNN-created Sketch}
The students who believed that humans created it:
\begin{quote}
\textit{    It shows the transformation of a drawing in a way that I think only a human could do.
}\end{quote}

While most students thought it was generated by AI:

\begin{quote}
\textit{    The progression...through tiny detail changes, reminds me of how machine learning adapts through generations and slowly changes behaviour.
}\end{quote}

\subsubsection{Q3. MusicVAE-generated Melody}
Most students believed it's created by AI:

\begin{quote}
\textit{It seems very emotionless and the melody has a very similar sound all throughout all of its duration
}\end{quote}



Some disagreed or were unsure:
\begin{quote}
\textit{ I feel like it could be both, but I'm gonna say human because...an AI would do more patterned work.
}\end{quote}

\subsection{Quiz}

The quiz was conducted in the ``Loss \& Accuracy!" activity of Module 3. 22 valid responses were collected, the accuracies from Q1 to Q5 are 90.9 \%, 68.2 \%, 95.5 \%, 68.2 \%, and 59.1 \% (Table~\ref{table:1}). Q1 and Q3 received the highest accuracies and were both related to inputs to the VAEs; the three questions that received lower accuracies were all about the structure and roles of VAEs. The results suggested that students could easily tell the inputs to an AI while struggling with the detailed structures of a specific model. Q5 received the lowest accuracy, with 6 (out of 22) responses on either encoding or decoding and 3 unanswered probably because of confusion: MusicVAE's 2D palette indeed involves all three parts of a VAE, although the ``drawing path" in specifically meant interpolation, which only 59.1 \% of the students selected correctly.

\begin{table}[htb!]
\small 
\centering
\renewcommand\cellgape{\Gape[1.5mm][1.5mm]} 
\begin{tabular} { | c | c | }
  \hline
  Questions & Options \\
  \hline
  \makecell[l]{Q1. What kind of\\ data was SketchRNN\\ learning from?} & 
 \makecell[l]{\textbf{A. Human Sketches [90.9\%]} \\ B. AI Sketches \\ C. Animal Pictures [4.5\%] \\ D. Nothing! It's magic! }\\
  \hline
  \makecell[l]{Q2. Where does \\ the interpolation \\ happen?} & 
  \makecell[l]{A. Encoder \\ \textbf{B. z (latent space) [68.2\%]} \\ C. Decoder [22.7\%] \\ D. What is the interpolation again?}\\
  \hline
  \makecell[l]{Q3. What kind of\\data was MusicVAE\\ learning from?} & \makecell[l]{\textbf{A. Melodies/Beats [95.5\%]} \\ B. Interpolations [4.5\%] \\ C. I don't know! \\ D. Pictures} \\ 
  \hline
  \makecell[l]{Q4. Where does \\ the MusicVAE's \\ palette reside?} &
  \makecell[l]{A. It's just a sketch! \\ B. Encoder [9.1\%] \\ C. Decoder [22.7\%] \\ \textbf{D. z (latent space) [68.2\%]}} \\ 
  \hline
  \makecell[l]{Q5. What does \\the drawing path\\ actually mean?} & 
  \makecell[l]{A. Encoding [9.1\%] \\ \textbf{B. Interpolations on z [59.1\%]} \\ C. Decoding [18.2\%] \\ D. Come on...why this picture again?}\\ 
 \hline
\end{tabular}
\caption{Quiz questions and responses (N=22). Bold options are correct, options without percentages indicate no responses. Besides, there were 4.5\%, 9.1\%, 0\%, 0\%, and 13.6\% of students who did not answer Q1-5, respectively.}
\label{table:1}
\end{table}


\subsection{Post Completion Survey}
\label{feebackSection}

\begin{figure}[t]
\centering
\includegraphics[width=\columnwidth]{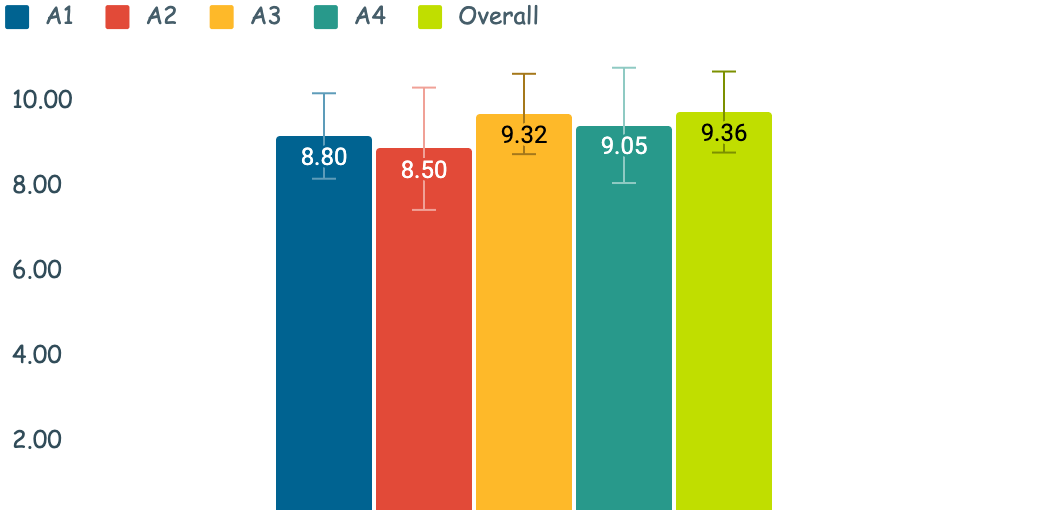}
\caption{Average ratings of four key activities and the lesson overall (0-10 scale, error bars report standard deviations)}
\label{feedback}
\end{figure}

Students provided feedback in a post completion survey in the ``Our VAEs Journey" activity of Module 3, where students were asked to rate four key activities: (1) Shadows and Plato's Cave, (2) Shadow Matching Game, (3) Digits Interpolation Notebook, and (4) VAEs and Arts. Students were also required to provide ratings and feedback for the lesson overall. 20 valid responses are collected for (1) and (4); 22 were collected for (2), (3), and overall. (Figure~\ref{feedback})


Overall, the lesson received a high 9.36 average ratings, and all four activities received average ratings above 8.5. In fact, a student said:

\begin{quote}
\textit{I really enjoyed this session about VAE's and learned a lot from it. I think that the tools and examples that we used were very engaging and helped understand the basic concepts of a VAE, and I think that the lesson did not really need any improvement. }


\end{quote}

The Shadow Matching Game received the lowest scores with 8.5 on average, and the Digits Interpolation Notebook received the highest with 9.32 on average. This suggests that students were having troubles with the game:

\begin{quote}
\textit{ If the 3d image activity [Shadow Matching Game] interface could be made a little simpler then that would make it way easier but the activity overall was great.}
\end{quote}
Most students preferred the notebook activity and even urged for diving deeper into the actual Python code, which we avoided on purpose to simplify the lesson:
\begin{quote}
\textit{- Maybe even explain the code aspect of training the VAEs, but its understandable that that is not done as Python would be a requirement for that. 
}


\textit{- It could be interesting to teach about the code behind the VAE and how that works.
}
\end{quote}

We also noticed that those with little or no coding experience struggled with the current notebook activity and the Google Colab interface:

\begin{quote}
\textit{The digits activity too was great but I did not understand the interface completely, so if that could be made simpler. It was a little bit complex and difficult to understand at some parts. 
}
\end{quote}


Students gave ``VAEs and Arts?" an average of 9.05 while gave ``Shadows and Plato's Cave" an average of 8.80. This suggested that students might prefer hands-on activities that allowed them to train or explore AI (A3\&A4), while not entirely enjoying the concept explanation activities (A1\&A2): 

\begin{quote}
\textit{- I really liked this lesson because it was able to explain the basics and dive into applications almost immediately, which truly helped me to learn because I was able to see how to do it myself.
}


\textit{- Plato's cave could be given at the end so that we can relate our knowledge to the theory. When it was introduced at the beginning, it was a little bit confusing. 
}

\textit{- Activities were engaging as each one applied what we had just learned. It was cool to see the application of each aspect of the VAE.}

\end{quote}


Students demanded more real-world applications:

\begin{quote}
\textit{- It would also be interesting to learn about how people are using this in the industry.}

\textit{- Maybe in the near future you could try adding more examples to your course to do with vae with a wider variation that is open to real life examples.}

\end{quote}


However, since this lesson focuses on VAE alone,  it lacks the bigger picture of AI. Furthermore, due to the time constraint, it doesn't provide an in-depth explanation of concepts such as neural networks, encoder, decoder, etc., which were suggested by the students as well:

\begin{quote}






\textit{- Solid definitions for VAEs, encoders, latent space, and decoders in the beginning would be helpful.}

\textit{- There could be a bigger explanation of neural networks and how those work more precisely.}


\textit{- Introducing what AI is and what the VAE is would help students better understand. Also explaining the encoder, decoder, and z space in a more broader perspective would be helpful (such as providing a definition and what each one does in a general perspective rather than learning just from examples).
}

\end{quote}

\section{Discussion}
From the pre-assessment, we see students came into class with good understandings of what is created by AI, especially for music, as melodies generated by MusicVAE sound emotionless and artificial. However, some were unsure since both the sketches and the melodies are more creative than the AI-generated work they had seen before, and thus curious about how AI achieved that. We attempted to satiate that curiosity by teaching this VAE lesson. We saw from the quiz that after the lesson, most students could tell the roles of the encoder, decoder, and latent space and understood how VAEs perform interpolation to create the art pieces they saw in the pre-assessment. Overall, this lesson was successful, as students gave high ratings for all four key activities and positive feedback for the lesson as a whole. However, in attempting to design the lessons for no pre-requisite knowledge in mathematics, statistics, and coding, students with relative experience demanded more challenging materials, such as working on the Python code of the digits interpolation notebook. Those without backgrounds in CS expressed confusion about the definitions of the encoder and decoder and the bigger picture of AI. Students were also confused about the role of the decoder and latent space in real-world applications. For example, the 2D palette for Beat Blender was always recognized as a decoder for students, instead of ``a latent space, that used encoder and decoder to perform interpolation". Students also seemed to have struggled with the lack of familiarity with the Colab interface. While great collaborative programming tools such as Google Colab, Jupyter Notebooks, etc., have been used as effective teaching aids for undergraduate students, they seem too complex for K-12 students, even with simplified UI.

In summary, we suggest:
(1) Teachers should adapt this lesson according to students' backgrounds. We have prepared that by letting teachers use the \textit{Easy} or \textit{Hard} mode in shadow matching game and allowing teachers to expose the code underneath the notebook; (2) For those without AI or CS background, this lesson is better to be taught with a bigger Introduction to AI curriculum to provide the foundational concepts, such as neural networks, training and testing, etc. that is needed for learning VAEs; (3) Students enjoyed interactive applications of VAEs, so including more hands-on activities as we did with the digits interpolation notebook and tools of SketchRNN and MusicVAE are helpful; (4) For those working on Creative AI or AI and Philosophy curricula, this lesson can be incorporated to enrich that; (5) Designers of educational technology tools can work towards developing a more beginner-friendly collaborative programming tool for K-12 students. 


\section{Limitations and Future Work}
\label{limitation}
(1) In the Digits Interpolation Notebook, we let the algorithm re-train the entire VAE. This approach makes the model overfit the custom digits that the students drew, but the results are much clearer for illustration. A better approach is to re-train only the last few layers, which is a widely used technique~\cite{bozinovski2020reminder, pan2010a,bozinovskiinfluence}; (2) For simplicity purposes, we did not dive deeper into the probability and distribution thus did not distinguish between VAE and AE. However, in the ``VAE and AE" section, we have shown that the Shadow Matching Game can be developed into a version with probability; (3) We only deployed the web version of the Shadow Matching Game due to limited access to the VR headsets and the restriction during the pandemic. Since the users lost one dimension on the 2D screen, they found it hard to manipulate the objects. We look forward to deploying the VR version for an immersive, interactive learning experience and potentially including philosophical thought experiments about reality such as brain-in-a-vat~\cite{horgan2011phenomental}; (4) The pilot sample might suffer from selection bias since those who replied to our call were more likely to have backgrounds and interests in coding and AI. In the future, we plan to conduct a full study in school classrooms, which would have more variation in students' capabilities and interests; (5) While our lesson discusses the constitution and applications of VAEs, there is a lack of reflection on the ethical implications of the technology, which has been included in previous AI for K-12 curricula~\cite{ali2021exploring, ali2019constructionism, lee2021developing, williams2021teacher} and could be added to our lesson to help students make informed decisions guided by morals and values; (6) We used an ancient and somewhat abstract philosophical allegory as a metaphor to disambiguate the concepts behind VAEs. We received expert feedback that it may be a far-fetched analogy and it would be appropriate to use more direct examples of internal representations such as concepts and symbols to get the idea across. Meanwhile, future work connecting AI concepts with modern philosophical ideas needs to be conducted to explore the relationship between AI literacy and philosophical literacy. For instance, the study of the nature of art, what AI considers art or not, and how historical biases and inherent aesthetics play a role. 

\section{Conclusion}
This paper presents a VAEs lesson\footnote{All resources are available at {\textbf{{https://raise.mit.edu/vae/}}}} designed for high school students. A user study with 22 students shows that this lesson successfully provided basic understandings of VAEs through interactive tools, a philosophical metaphor, and artistic explorations. The lesson can be used independently to teach students about a key generative AI concept, in conjunction with existing Creative AI curricula~\cite{ ali2021exploring, ali2019constructionism}, or in other AI and Philosophy curricula. We hope the K-12 AI education community can use our findings and resources to provide meaningful VAEs lessons.



\section{Acknowledgments}
We thank the EAAI reviewers and members of the Personal Robots Group, MIT Media Lab, for their feedback. We also thank all students and parents for their insights and participation in the program.
\bibliography{vae}
\end{document}